\documentclass[3p,times]{elsarticle}

\usepackage{hyperref}
\usepackage{txfonts}
\usepackage{graphicx, subfigure}
\usepackage{adjustbox,lipsum}
\usepackage{tabularx}
\usepackage{dcolumn}
\usepackage{lineno}
\usepackage{tabularx}
\biboptions{sort&compress}
\usepackage{color}

\usepackage{amssymb}

\makeatletter
\def\ps@pprintTitle{%
 \let\@oddhead\@empty
 \let\@evenhead\@empty
 \def\@oddfoot{}%
 \let\@evenfoot\@oddfoot}
\makeatother

\begin{document}

\begin{frontmatter}



\title{Van der Waals density-functional theory study for bulk solids with BCC, FCC, and diamond structures}


\author[sejongaddress]{Jinwoo Park}

\author[uosaddress]{Byung Deok Yu\corref{cor}}
\ead{ybd@uos.ac.kr}

\author[sejongaddress]{Suklyun Hong\corref{cor}}
\ead{hong@sejong.ac.kr}

\cortext[cor]{Corresponding authors}
\address[sejongaddress]{Graphene Research Institute and Department of Physics, Sejong University, Seoul 143-747, Republic of Korea}
\address[uosaddress]{Department of Physics, University of Seoul, Seoul 130-743, Republic of Korea}

\begin{abstract}
Proper inclusion of van der Waals (vdW) interactions in theoretical simulations based on standard density functional theory (DFT) is crucial to describe the physics and chemistry of systems such as organic and layered materials. Many encouraging approaches have been proposed to combine vdW interactions with standard approximate DFT calculations. Despite many vdW studies, there is no consensus on the reliability of vdW methods. To help further development of vdW methods, we have assessed various vdW functionals through the calculation of structural properties at equilibrium, such as lattice constants, bulk moduli, and cohesive energies, for bulk solids, including alkali, alkali-earth, and transition metals, with BCC, FCC, and diamond structures as the ground state structure. These results provide important information for the vdW-related materials research, which is essential for designing and optimizing materials systems for desired physical and chemical properties.
\end{abstract}


\end{frontmatter}


\section{Introduction}
\label{sec1}
The development of approximate density functional theory (DFT) \cite{Kohn1964,Kohn1965} methods that are able to account for van der Waals (vdW) interactions has attracted a great deal of interest due to its importance in the theoretical description of organic or layered materials as well as that of physical, chemical, and biological processes~\cite{intro1,intro2,intro3,intro4,intro5,RPA1,Haas}. Many encouraging schemes and algorithms have been proposed to include vdW interactions in theoretical simulations based on standard DFT. One of the promising vdW approaches is the vdW density functional (vdW-DF) method, which does not depend on external input parameters and is based directly on the electron density~\cite{llvdw1}. In the vdW-DF method, the exchange-correlation (XC) energy is given as
\begin{equation}
E_{\rm XC} = E_{\rm X}^{\rm GGA} + E_{\rm C}^{\rm LDA} + E_{\rm C}^{\rm nl},
\end{equation}
where $E_{\rm X}^{\rm GGA}$ is the generalized gradient approximation (GGA) to the exchange energy, $E_{\rm C}^{\rm LDA}$ is the local density approximation (LDA) to the correlation energy, and $E_{\rm C}^{\rm nl}$ is the nonlocal electron correlation energy. In the case of the vdW-DF approach, the computational time increases by 50\%{} compared to standard DFT calculations such as the LDA and the GGA calculations~\cite{timespeed}. Depending on the selection of the exchange functional, there are many vdW-DF methods. Here we consider the five vdW-DF functionals; revPBE-vdW~\cite{llvdw1}, rPW86-vdW2~\cite{llvdw2,llvdw3}, optPBE-vdW \cite{optPBE1}, optB88-vdW~\cite{optPBE1}, and optB86b-vdW~\cite{optPBE2}. In addition, there is another widely used vdW approach, the so-called dispersion-corrected DFT-D in which an atom-pairwise potential is added to a standard DFT result. In the original DFT-D scheme \cite{Grimme2004}, the predetermined constant dispersion coefficients are assigned to an element irrespective of its environment. To improve this, the dispersion coefficients are further modified to vary with the environment of an element. In contrast to the vdW-DF schemes, the DFT-D schemes do not add a significant computational cost compared to the standard DFT calculations. In the DFT-D schemes, we consider the five vdW functionals; DFT-D2~\cite{DFTD2}, DFT-D3~\cite{DFTD3}, DFT-D3(BJ)~\cite{DFTD3BJ}, DFT-TS~\cite{DFTTS}, and DFT-TS-SCS~\cite{DFTTSSCS1,DFTTSSCS2}. The environment-dependent DFT-D3 scheme has zero damping for small interatomic distances, whereas the DFT-D3(BJ) scheme has rational damping to finite values (BJ-damping) as Becke and Johnson proposed. Grimme {\em et al.} suggested that the DFT-D3(BJ) performed slightly better than the DFT-D3 for noncovalently-bonded materials systems~\cite{DFTD3BJ}. In the DFT-TS scheme, the dispersion coefficients are determined by employing the partitioning of the electron density~\cite{DFTTS}. The DFT-TS scheme can be further modified by incorporating self-consistent long-range screening effects \cite{DFTTSSCS1,DFTTSSCS2}. This modified scheme is herein called the DFT-TS-SCS functional. Tkatchenko {\em et al.} report that the DFT-TS-SCS functional performs better than the DFT-TS functional~\cite{DFTTSSCS1,DFTTSSCS2}. However, despite many vdW studies, the assessment of the performance of the vdW functionals on a broad range of material systems is lacking.

In the present work, we have investigated structural properties (lattice constants, bulk moduli, and cohesive energies) at equilibrium for bulk solids with body centered cubic (BCC), face centered cubic (FCC), and diamond (DIA) structures, to assess the performance of various vdW functionals based on the DFT. We herein consider the ten vdW functionals implemented in the Vienna {\em Ab-initio} Simulation Package (VASP) code~\cite{VASP1,VASP2,VASP3,VASP4}; revPBE-vdW, rPW86-vdW2, optPBE-vdW, optB88-vdW, optB86b-vdW, DFT-D2, DFT-D3, DFT-D3(BJ), DFT-TS, and DFT-TS-SCS functionals. For comparison, the LDA and GGA calculations were also performed. Our calculations show that the five vdW functionals of optB86b-vdW, optB88-vdW, optPBE-vdW, DFT-D3, and DFT-D3(BJ) give better performance compared to the other vdW functionals. Differences among the results from the five vdW functionals are also discussed. These results provide important information for further development of vdW methods to improve the description of a wide range of materials systems.

The paper is organized as follows. In Sec. \ref{Calculation}, the computational method and settings used in this study are briefly described. The results and discussion are presented in Sec. \ref{Results}. Finally, the conclusions are stated in Sec. \ref{Summary}.

\section{Computation method}
\label{Calculation}
We employed the VASP code to perform the DFT calculations, including spin effects for magnetic elements, with various vdW functionals \cite{VASP1,VASP2,VASP3,VASP4}. In this work, we considered ten vdW functionals implemented in the VASP; revPBE-vdW \cite{llvdw1}, rPW86-vdW2 \cite{llvdw2,llvdw3}, optPBE-vdW~\cite{optPBE1}, optB88-vdW \cite{optPBE1}, optB86b-vdW~\cite{optPBE2}, DFT-D2 \cite{DFTD2}, DFT-D3 \cite{DFTD3}, DFT-D3(BJ) of Becke-Jonson \cite{DFTD3BJ}, DFT-TS \cite{DFTTS}, and DFT-TS-SCS \cite{DFTTSSCS1,DFTTSSCS2}. For comparison, LDA and GGA calculations were also performed using the Ceperley-Alder~\cite{LDA} and the Perdew-Burke-Ernzerhof (PBE)~\cite{GGA} expressions, respectively, for the exchange-correlation functional. In the case of the DFT-D schemes [DFT-D2, DFT-D3, DFT-D3(BJ), DFT-TS, and DFT-TS-SCS], we used the PBE parameterization of the GGA for the exchange-correlational functional. For electron-ion interactions, the projector augmented-wave (PAW) method \cite{PAW1,PAW2} was used.  We considered 29 elements with BCC, FCC, and diamond  structures as the ground state in the bulk phase. In the calculations, the electronic wave functions were expanded by plane waves with an energy cutoff of 700 eV. The {\bf k}-space integration was performed using a $\Gamma$-centered 12$\times$12$\times$12 mesh in the Brillouin zone (BZ) of the primitive cell. The tetrahedron method with Bl\"ochl corrections~\cite{tetra1,tetra2} was used to improve the computational convergence. We performed total energy calculations to obtain the ground state properties such as the equilibrium lattice constant, the bulk modulus, and the cohesive (atomization) energy. The ground state properties were determined by fitting the calculated total energy as a function of the volume to the Birch-Murnaghan equation of state \cite{Birch1947,Jang2011,Jang2012}. In the fitting, a set of eleven different volumes around the experimental equilibrium volume corresponding to the equilibrium lattice constant was used.

\section{Results and discussion}
\label{Results}
\subsection{Lattice constant}
\label{LatticeConstant}

\begin{table*}[!ht]
\caption{ Equilibrium lattice constants (in \AA) for bulk solids with BCC, FCC, and diamond (DIA) structures using the ten vdW functionals. In addition, standard DFT calculation results using LDA and GGA are also shown. The experimental lattice constants \cite{kittel} are measured at finite temperatures, in contrast to the theoretical lattice constants obtained from ground-state electronic structure calculations at zero temperature. For comparison, the experimental values were corrected to the $T=0$ limit using thermal expansion corrections \cite{csonka} for the solids (denoted with an asterisk) whose zero-point anharmonic expansion values are available. }
\label{table:ALAT}
\noindent \adjustbox{max width=\textwidth}{
		\begin{tabular}{ccccccccccccccc}
			\hline \hline
			Solid   & Crystal   & revPBE & rPW86 & optB86b & optB88 & optPBE & DFT & DFT & DFT      & DFT  & DFT        & PAW & PAW & Expt. \\
			            & structure & vdW      & vdW2   & vdW        & vdW      & vdW       & D2   & D3    & D3(BJ) & TS    & TS-SCS  & LDA & GGA &            \\
			\hline
Li	&	BCC	&	3.453 	&	3.393 	&	3.457 	&	3.435 	&	3.442 	&	3.270 	&	3.374 	&	3.329 	&	2.607 	&	 3.122 	&	3.367 	&	3.439 	&	3.449$^{\ast}$ 	\\
C	&	DIA	&	3.599 	&	3.605 	&	3.570 	&	3.575 	&	3.584 	&	3.564 	&	3.565 	&	3.558 	&	3.553 	&	 3.564 	&	3.535 	&	3.573 	&	3.543$^{\ast}$	\\
Na	&	BCC	&	4.214 	&	4.135 	&	4.176 	&	4.152 	&	4.178 	&	3.980 	&	4.161 	&	4.078 	&	4.131 	&	 3.852 	&	4.055 	&	4.193 	&	4.210$^{\ast}$	\\
Al	&	FCC	&	4.084 	&	4.087 	&	4.034 	&	4.054 	&	4.058 	&	4.010 	&	4.007 	&	3.982 	&	3.921 	&	 4.002 	&	3.984 	&	4.040 	&	4.020$^{\ast}$	\\
Si	&	DIA	&	5.513 	&	5.540 	&	5.456 	&	5.469 	&	5.484 	&	5.412 	&	5.453 	&	5.420 	&	5.446 	&	 5.438 	&	5.403 	&	5.469 	&	5.416$^{\ast}$	\\
K	&	BCC	&	5.289 	&	5.174 	&	5.198 	&	5.162 	&	5.222 	&	5.156 	&	5.226 	&	5.163 	&	4.345 	&	 4.943 	&	5.043 	&	5.284 	&	5.212$^{\ast}$	\\
Ca	&	FCC	&	5.553 	&	5.493 	&	5.464 	&	5.450 	&	5.501 	&	5.381 	&	5.498 	&	5.441 	&	5.169 	&	 5.301 	&	5.338 	&	5.532 	&	5.553$^{\ast}$	\\
V	&	BCC	&	3.007 	&	3.026 	&	2.958 	&	2.969 	&	2.982 	&	2.970 	&	2.938 	&	2.932 	&	2.917 	&	 2.954 	&	2.912 	&	2.978 	&	3.030 	\\
Cr	&	BCC	&	2.866 	&	2.887 	&	2.821 	&	2.833 	&	2.843 	&	2.815 	&	2.809 	&	2.809 	&	2.770 	&	 2.808 	&	2.779 	&	2.836 	&	2.880 	\\
Fe	&	BCC	&	2.873 	&	2.889 	&	2.806 	&	2.821 	&	2.838 	&	2.802 	&	2.805 	&	2.806 	&	2.779 	&	 2.812 	&	2.749 	&	2.832 	&	2.870 	\\
Ni	&	FCC	&	3.572 	&	3.609 	&	3.488 	&	3.511 	&	3.529 	&	3.459 	&	3.476 	&	3.477 	&	3.424 	&	 3.489 	&	3.422 	&	3.518 	&	3.520 	\\
Cu	&	FCC	&	3.705 	&	3.750 	&	3.601 	&	3.629 	&	3.651 	&	3.571 	&	3.568 	&	3.568 	&	3.547 	&	 3.607 	&	3.524 	&	3.635 	&	3.595$^{\ast}$	\\
Ge	&	DIA	&	5.895 	&	5.973 	&	5.764 	&	5.798 	&	5.828 	&	5.666 	&	5.760 	&	5.723 	&	5.749 	&	 5.730 	&	5.647 	&	5.783 	&	5.640$^{\ast}$	\\
Rb	&	BCC	&	5.666 	&	5.545 	&	5.535 	&	5.501 	&	5.578 	&	5.443 	&	5.615 	&	5.551 	&	5.666 	&	 5.666 	&	5.374 	&	5.668 	&	5.576$^{\ast}$	\\
Sr	&	FCC	&	6.057 	&	6.004 	&	5.921 	&	5.911 	&	5.980 	&	5.760 	&	5.985 	&	5.924 	&	6.306 	&	 5.731 	&	5.788 	&	6.026 	&	6.045$^{\ast}$	\\
Nb	&	BCC	&	3.340 	&	3.377 	&	3.288 	&	3.303 	&	3.314 	&	3.308 	&	3.265 	&	3.264 	&	3.214 	&	 3.269 	&	3.247 	&	3.308 	&	3.300 	\\
Mo	&	BCC	&	3.183 	&	3.219 	&	3.139 	&	3.154 	&	3.161 	&	3.145 	&	3.124 	&	3.122 	&	3.084 	&	 3.123 	&	3.106 	&	3.151 	&	3.150 	\\
Rh	&	FCC	&	3.879 	&	3.942 	&	3.803 	&	3.829 	&	3.841 	&	3.773 	&	3.786 	&	3.788 	&	3.767 	&	 3.806 	&	3.752 	&	3.824 	&	3.793$^{\ast}$	\\
Pd	&	FCC	&	4.012 	&	4.083 	&	3.905 	&	3.938 	&	3.957 	&	3.888 	&	3.886 	&	3.889 	&	3.912 	&	 3.921 	&	3.841 	&	3.942 	&	3.875$^{\ast}$	\\
Ag	&	FCC	&	4.242 	&	4.313 	&	4.091 	&	4.130 	&	4.163 	&	4.131 	&	4.073 	&	4.072 	&	4.068 	&	 4.117 	&	4.003 	&	4.147 	&	4.056$^{\ast}$	\\
Sn 	&	DIA	&	6.764 	&	6.876 	&	6.593 	&	6.637 	&	6.677 	&	6.505 	&	6.612 	&	6.571 	&	6.538 	&	 6.577 	&	6.478 	&	6.652 	&	6.490 	\\
Cs	&	BCC	&	6.134 	&	5.989 	&	5.942 	&	5.899	&	6.014 	&	4.235 	&	6.111 	&	6.039 	&	4.382 	&	 5.809 	&	5.760 	&	6.161 	&	6.039$^{\ast}$	\\
Ba	&	BCC	&	5.073 	&	5.057 	&	4.904 	&	4.915 	&	4.986 	&	4.203 	&	4.974 	&	4.935 	&	5.021 	&	 4.724 	&	4.768 	&	5.030 	&	4.995$^{\ast}$	\\
Ta	&	BCC	&	3.340 	&	3.375 	&	3.292 	&	3.306 	&	3.317 	&	3.260 	&	3.273 	&	3.276 	&	3.235 	&	 3.281 	&	3.248 	&	3.309 	&	3.300 	\\
W	&	BCC	&	3.203 	&	3.239 	&	3.163 	&	3.178 	&	3.184 	&	3.106 	&	3.147 	&	3.148 	&	3.123 	&	 3.154 	&	3.129 	&	3.172 	&	3.160 	\\
Ir	&	FCC	&	3.923 	&	3.986 	&	3.861 	&	3.886 	&	3.892 	&	3.767 	&	3.838 	&	3.843 	&	3.844 	&	 3.862 	&	3.819 	&	3.873 	&	3.840 	\\
Pt	&	FCC	&	4.032 	&	4.110 	&	3.949 	&	3.980 	&	3.990 	&	3.839 	&	3.918 	&	3.926 	&	3.934 	&	 3.952 	&	3.897 	&	3.968 	&	3.920 	\\
Au	&	FCC	&	4.245 	&	4.338 	&	4.122 	&	4.161 	&	4.181 	&	3.996 	&	4.099 	&	4.101 	&	4.113 	&	 4.134 	&	4.052 	&	4.157 	&	4.080 	\\
Pb	&	FCC	&	5.134 	&	5.233 	&	4.972 	&	5.018 	&	5.053 	&	5.072 	&	4.971 	&	4.948 	&	4.949 	&	 4.997 	&	4.875 	&	5.031 	&	4.902$^{\ast}$	\\
			\hline
		\end{tabular}
}
\end{table*}


\begin{figure*}[ht!]
     \begin{center}
        \subfigure{
            \label{fig1a}
            \includegraphics[width=0.47\textwidth]{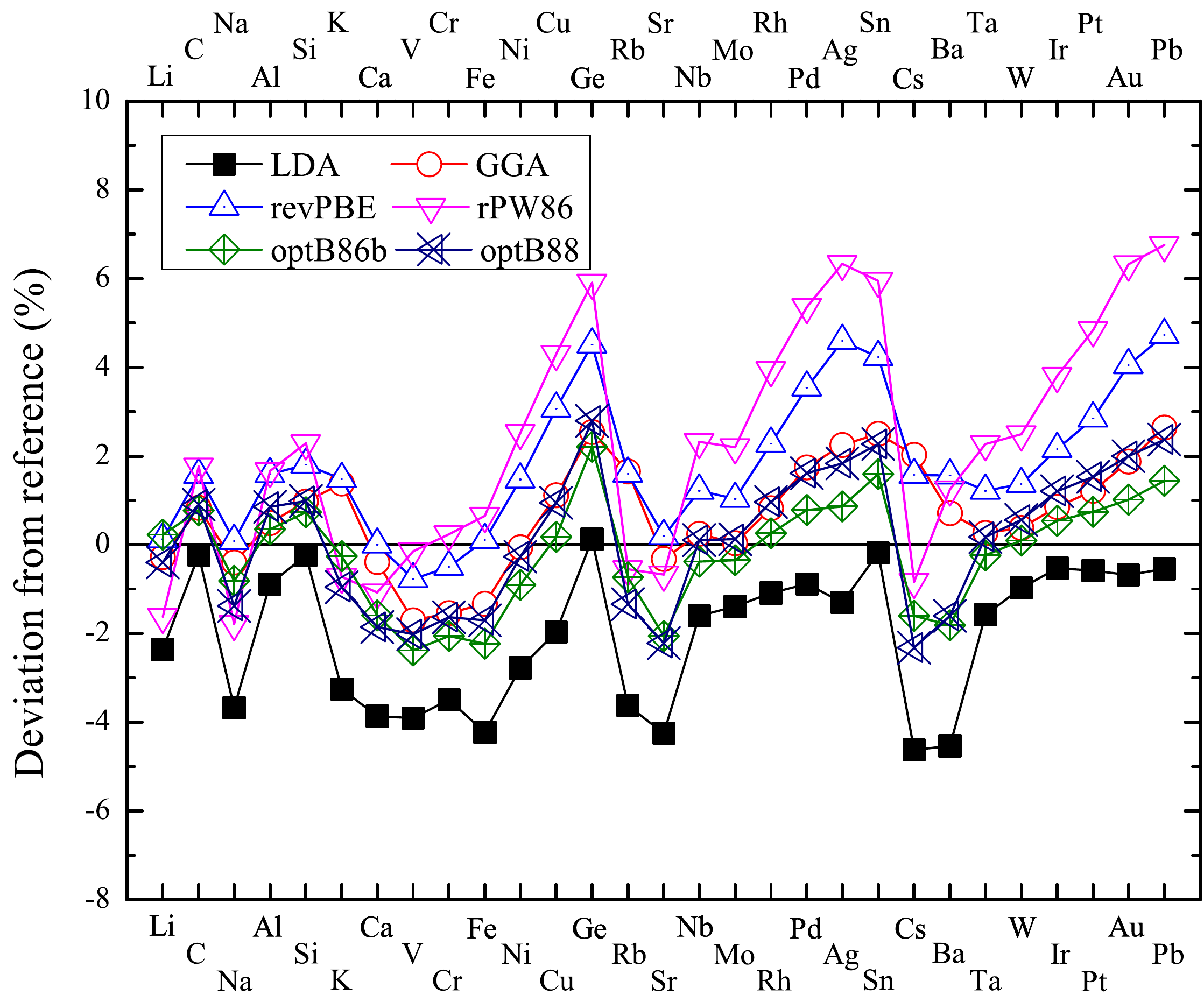}
        }
        \subfigure{
           \label{fig1b}
           \includegraphics[width=0.47\textwidth]{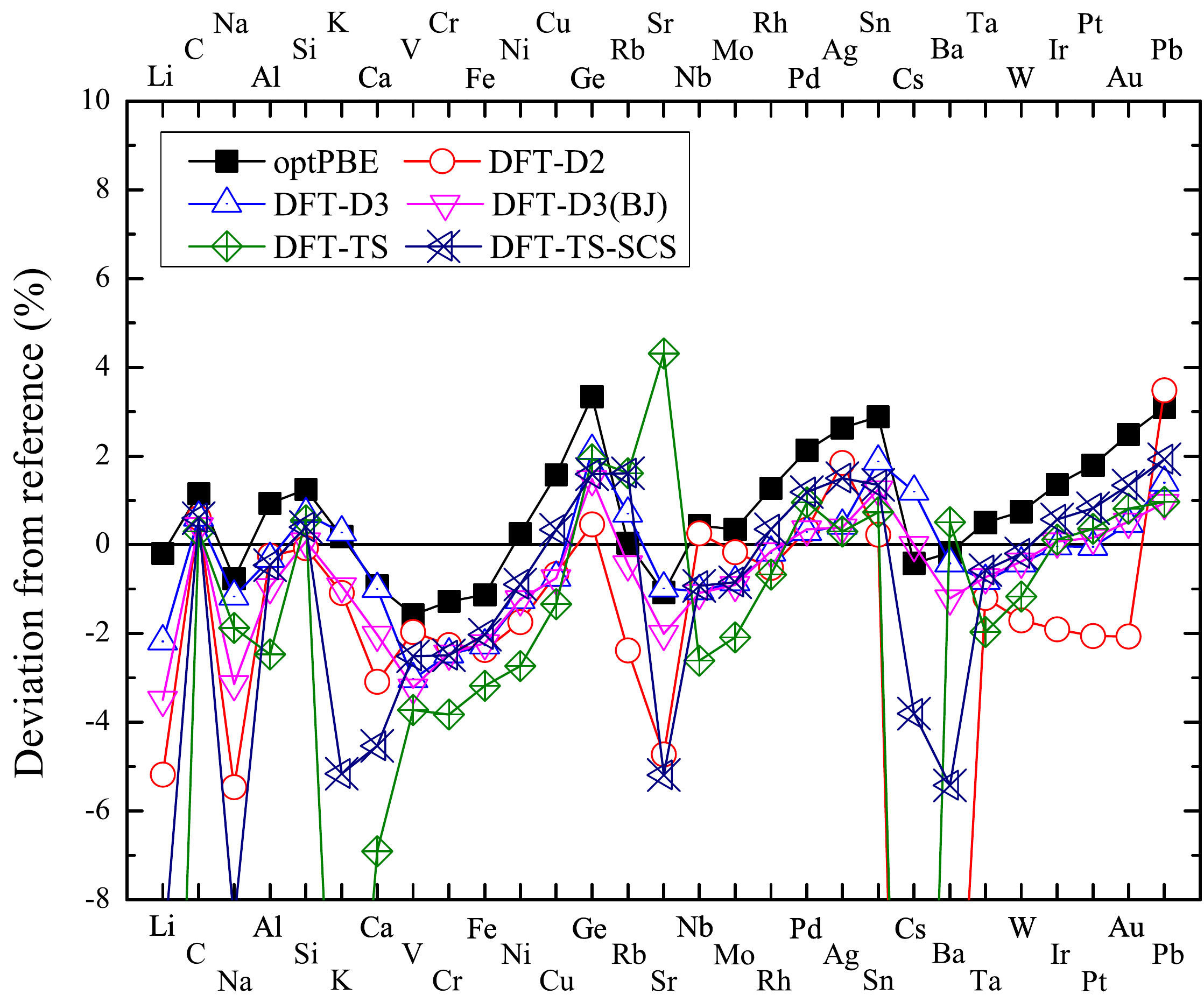}
        }
    \end{center}
    \caption{ Relative errors in the calculated equilibrium lattice constants with respect to the experimental values. Results are shown for the ten vdW functionals and the standard DFT functionals of LDA and GGA. The positive (negative) values in the relative errors represent the larger (smaller) lattice constants than the experimental values. }
    \label{fig1}
\end{figure*}

The equilibrium lattice constants calculated with the ten vdW functionals are summarized in Table~\ref{table:ALAT}, and the relative errors in the equilibrium lattice constants with respect to the experimental values are shown in Fig.~\ref{fig1}. For comparison, we also present the results of the standard DFT functionals of LDA and GGA. The standard GGA functional gives the relative errors in the range of $\pm 2\%$, while the standard DFT functional of LDA shows shorter equilibrium lattice constants than those from the other functionals, indicating the well known overbinding of atoms in the LDA approach \cite{LDAWEAKPOINT,LDAoverbinding}. In the case of the vdW functionals, optB86b-vdW, optB88-vdW, optPBE-vdW, and DFT-D3 show the relative errors in the range of $\pm 3\%$. The DFT-D3(BJ) functional shows the relative errors in the range from $-3\%$ to $+1\%$ for all elements except for Li. The DFT-TS and DFT-TS-SCS functionals give results comparable to the other vdW functionals, except for the alkali (Li, Na, K, Cs) and alkali-earth (Ca, Sr, Ba) metals. For these two vdW functionals, the DFT-TS-SCS scheme with the self-consistent screening (SCS) effects shows better performance than the DFT-TS scheme without the SCS effects \cite{DFTTSSCS1,DFTTSSCS2}. In the case of revPBE-vdW and rPW86-vdW2, the relative errors range from $-3\%$ to $+6\%$. The relative errors are observed to be more scattered compared to those from the vdW functionals of optPBE-vdW, optB88-vdW, optB86b-vdW, DFT-D3, and DFT-D3(BJ). This behavior becomes more significant with the increase of the atomic number.

\begin{figure*}[ht!]
     \begin{center}
        \subfigure{%
            \label{fig2a}
            \includegraphics[width=0.47\textwidth]{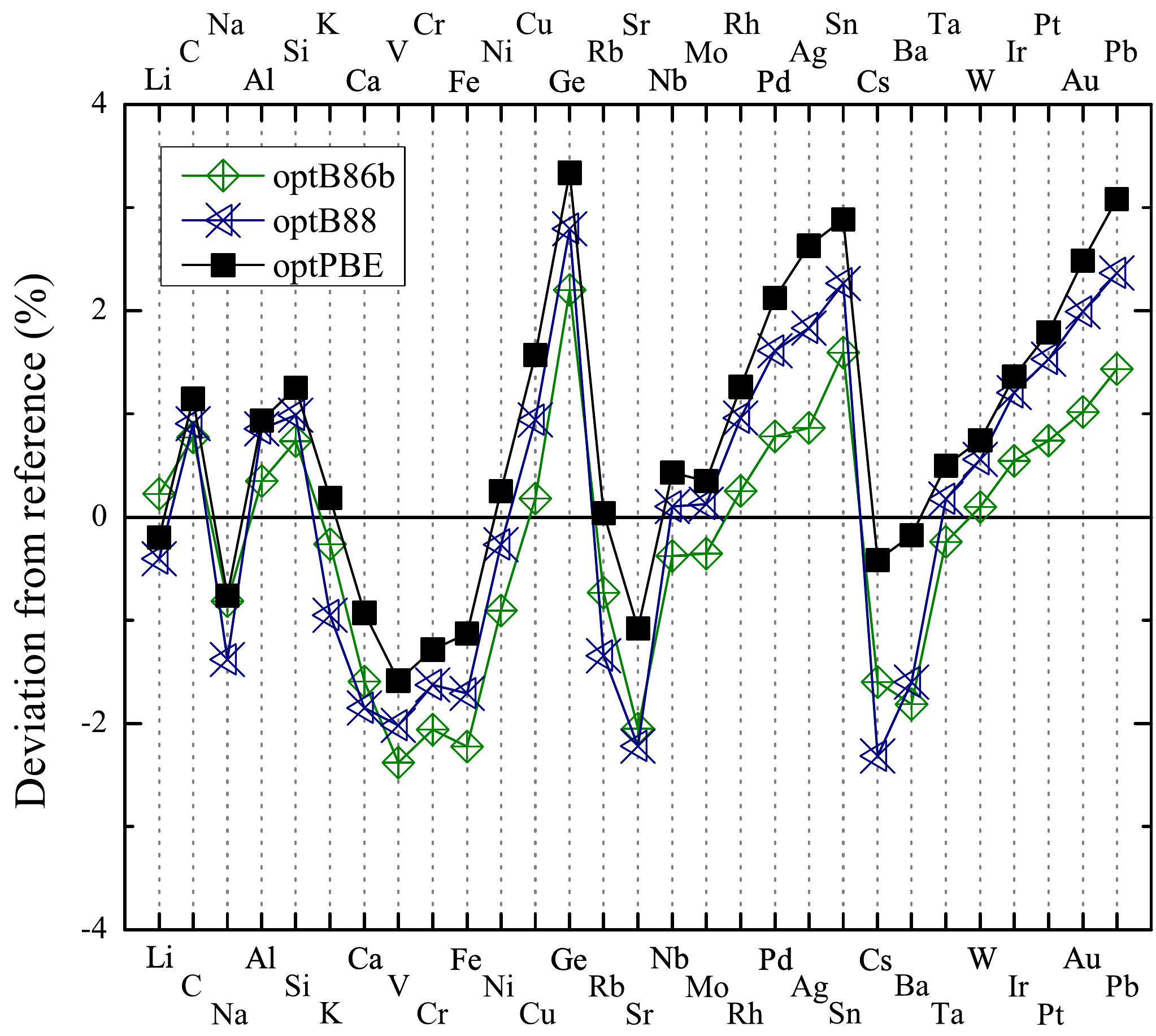}
        }%
        \subfigure{%
           \label{fig2b}
           \includegraphics[width=0.47\textwidth]{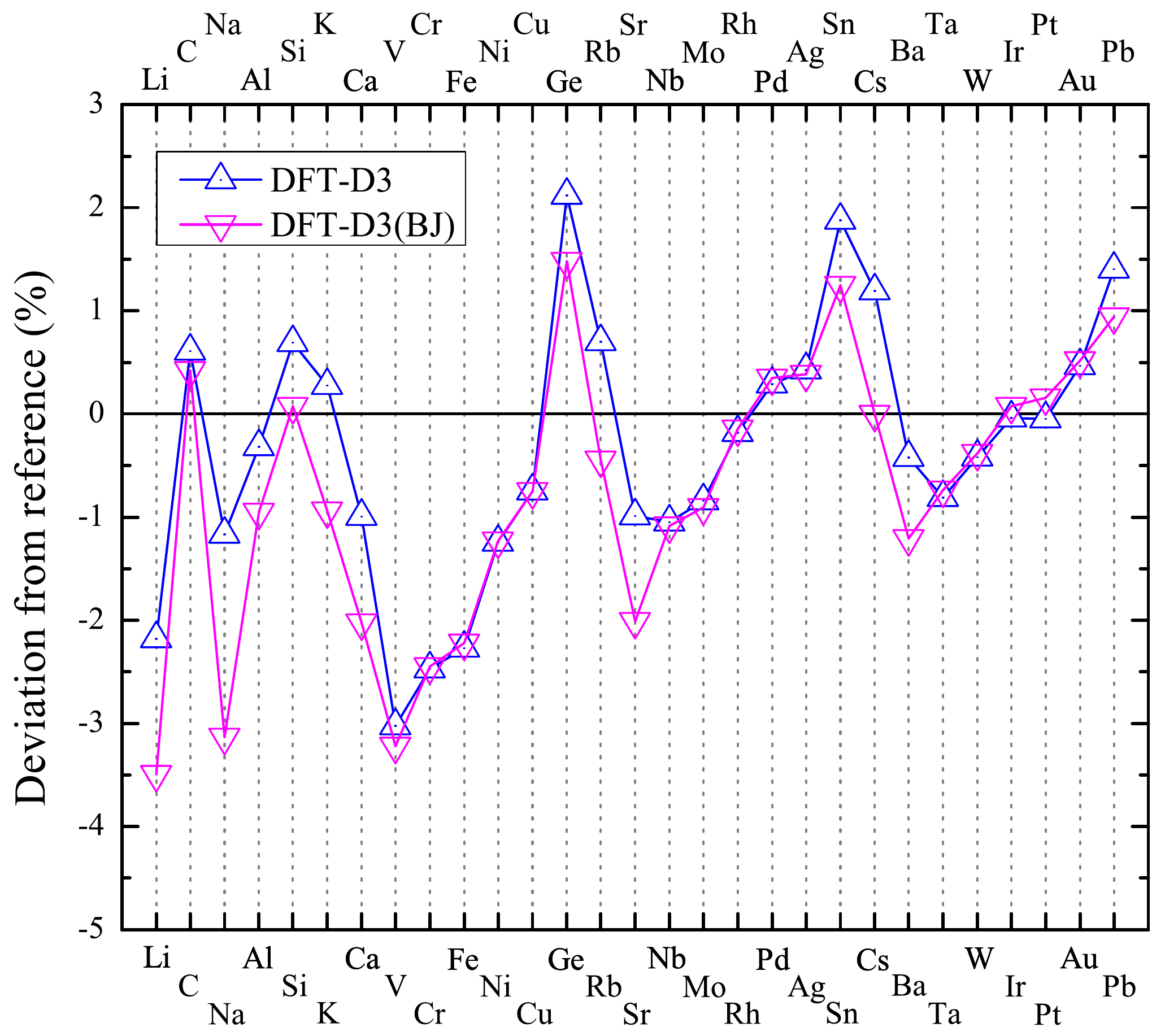}
        }
    \end{center}
    \caption{ Comparison of the relative errors in the calculated equilibrium lattice constants with respect to the experimental values.}
    \label{fig2}
\end{figure*}

 To further aid our understanding, we discuss the differences among the five vdW results of optPBE-vdW, optB88-vdW, optB86b-vdW, DFT-D3, and DFT-D3(BJ). In the case of DFT-D3, and DFT-D3(BJ), Grimme {\em et\ al.} \cite{DFTD3BJ} reported that the DFT-D3(BJ) vdW functional with rational damping to finite values for small interatomic distances performed slightly better than the DFT-D3 functional with zero damping when used for noncovalently-bonded systems. Our calculations show that both schemes give very similar results (see Fig. \ref{fig2}). For alkali and alkali-earth metals, however, the DFT-D3 functional gives much better performance than the DFT-D3(BJ) functional (see Fig. \ref{fig2}). In the case of the vdW-DF functionals of optPBE-vdW, optB88-vdW, and optB86b-vdW, the optB86b-vdW functional gives either better or comparable performance in equilibrium lattice constants compared to the optPBE-vdW and optB88-vdW functionals (see Fig. \ref{fig2}).


\subsection{Bulk modulus}
\label{BulkModulus}

\begin{table*}[!ht]
\caption{Bulk moduli (in GPa) for bulk solids with BCC, FCC, and diamond (DIA) structures using the ten vdW functionals. In addition, standard DFT calculation results using LDA and GGA are also shown. The theoretical values are compared to the experimental data. The experimental bulk moduli \cite{kittel} were corrected to the $T=0$ limit using thermal effects and zero-point phonon effects (ZPPE) \cite{csonka} for the solids (denoted with an asterisk) whose ZPPE values are available. }
\label{table:BM}
\noindent \adjustbox{max width=\textwidth}{
		\begin{tabular}{ccccccccccccccc}
			\hline \hline
			Solid   & Crystal & revPBE & rPW86 & optB86b & optB88 & optPBE & DFT & DFT & DFT      & DFT  & DFT        & PAW & PAW & Expt. \\
			             & structure & vdW      & vdW2   & vdW        & vdW     & vdW      & D2   & D3    & D3(BJ)  & TS   & TS-SCS & LDA & GGA &           \\
						\hline

Li	&	BCC	&	13.694 	&	14.724 	&	13.365 	&	13.748 	&	13.845 	&	14.315 	&	13.513 	&	15.596 	&	187.631 	 &	24.117 	&	15.108 	&	13.885 	&	13.3$^{\ast}$         \\
C	&	DIA	&	404.522 	&	396.193 	&	432.393 	&	426.121 	&	418.944 	&	437.501 	&	433.770 	 &	439.529 	&	439.442 	&	434.991 	&	463.489 	&	430.649 	&	443.0 	\\
Na	&	BCC	&	7.558 	&	8.230 	&	7.845 	&	8.039 	&	7.917 	&	8.621 	&	8.244 	&	8.975 	&	14.229 	&	 14.219 	&	9.240 	&	7.826 	&	7.5$^{\ast}$	\\
Al	&	FCC	&	63.500 	&	57.305 	&	75.135 	&	67.233 	&	69.153 	&	68.368 	&	81.233 	&	85.247 	&	100.028 	 &	83.527 	&	82.568 	&	76.049 	&	79.4$^{\ast}$	\\
Si	&	DIA	&	81.566 	&	76.461 	&	89.842 	&	87.250 	&	85.647 	&	96.468 	&	88.212 	&	91.276 	&	90.158 	&	 93.356 	&	95.410 	&	87.711 	&	98.8    	\\
K	&	BCC	&	3.558 	&	3.947 	&	3.784 	&	3.936 	&	3.787 	&	3.761 	&	4.320 	&	4.132 	&	70.613 	&	 5.463 	&	4.470 	&	3.567 	&	3.7$^{\ast}$	\\
Ca	&	FCC	&	16.444 	&	16.833 	&	17.489 	&	17.138 	&	17.098 	&	17.105 	&	15.826 	&	17.963 	&	15.586 	&	 20.594 	&	18.463 	&	17.086 	&	18.4$^{\ast}$	\\
V	&	BCC	&	172.466 	&	169.806 	&	194.718 	&	188.971 	&	183.512 	&	173.953 	&	203.049 	 &	202.604 	&	235.407 	&	198.016 	&	213.713 	&	185.675 	&	161.9   	\\
Cr	&	BCC	&	237.579 	&	229.519 	&	272.448 	&	262.464 	&	254.717 	&	249.959 	&	279.851 	 &	277.020 	&	315.618 	&	283.531 	&	303.323 	&	259.661 	&	190.1  	\\
Fe	&	BCC	&	147.968 	&	154.123 	&	202.889 	&	192.267 	&	176.236 	&	177.166 	&	213.305 	 &	197.352 	&	242.121 	&	194.995 	&	247.046 	&	174.999 	&	168.3  	\\
Ni	&	FCC	&	165.299 	&	152.039 	&	208.651 	&	196.380 	&	186.375 	&	216.252 	&	207.194 	 &	208.723 	&	271.665 	&	211.216 	&	249.627 	&	193.570 	&	186.0  	\\
Cu	&	FCC	&	110.545 	&	99.879 	&	147.659 	&	137.055 	&	128.539 	&	146.506 	&	159.282 	&	 161.663 	&	168.634 	&	148.152 	&	182.498 	&	136.340 	&	137.0  	\\
Ge	&	DIA	&	48.428 	&	41.953 	&	60.226 	&	56.816 	&	54.096 	&	68.543 	&	59.596 	&	62.472 	&	64.857 	&	 65.049 	&	71.363 	&	58.379 	&	77.2    	\\
Rb	&	BCC	&	2.812 	&	3.119 	&	3.060 	&	3.192 	&	3.028 	&	3.004 	&	3.308 	&	3.143 	&	2.772 	&	 2.772 	&	3.570 	&	2.770 	&	2.9$^{\ast}$	\\
Sr	&	FCC	&	11.014 	&	12.123 	&	12.461 	&	12.744 	&	11.840 	&	12.945 	&	10.281 	&	12.195 	&	13.501 	&	 16.295 	&	14.315 	&	11.312 	&	12.4$^{\ast}$	\\
Nb	&	BCC	&	159.232 	&	153.666 	&	176.423 	&	171.920 	&	167.693 	&	166.781 	&	180.698 	 &	180.745 	&	210.983 	&	183.804 	&	189.481 	&	169.206 	&	170.2  	\\
Mo	&	BCC	&	244.495 	&	227.582 	&	276.592 	&	265.651 	&	259.989 	&	251.293 	&	283.879 	 &	283.494 	&	329.543 	&	288.041 	&	299.116 	&	265.888 	&	272.5  	\\
Rh	&	FCC	&	219.171 	&	191.033 	&	273.980 	&	255.757 	&	245.384 	&	267.614 	&	279.428 	 &	279.135 	&	289.154 	&	268.524 	&	314.511 	&	254.550 	&	270.4  	\\
Pd	&	FCC	&	133.434 	&	116.972 	&	183.377 	&	168.781 	&	157.180 	&	162.033 	&	187.936 	 &	187.733 	&	177.399 	&	173.192 	&	222.962 	&	162.581 	&	180.8  	\\
Ag	&	FCC	&	68.730 	&	61.864 	&	105.174 	&	95.643 	&	85.884 	&	64.765 	&	103.017 	&	108.873 	&	 111.938 	&	96.729 	&	135.435 	&	88.903 	&	100.7  	\\
Sn    &	DIA	&	30.578 	&	26.675 	&	38.707 	&	36.573 	&	34.487 	&	39.692 	&	38.626 	&	39.488 	&	44.543 	&	 39.704 	&	44.872 	&	35.827 	&	53.0             \\
Cs	&	BCC	&	2.019 	&	2.268 	&	2.173 	&	2.294 	&	2.167 	&	79.424 	&	2.254 	&	2.128 	&	82.969 	&	 2.921 	&	2.458 	&	1.937 	&	2.1$^{\ast}$	\\
Ba	&	BCC	&	8.906 	&	9.597 	&	9.595 	&	9.933 	&	9.349 	&	84.987 	&	8.264 	&	9.097 	&	9.425 	&	 11.487 	&	10.361 	&	8.768 	&	9.3$^{\ast}$	\\
Ta	&	BCC	&	187.810 	&	181.116 	&	205.534 	&	199.998 	&	196.462 	&	163.345 	&	212.634 	 &	209.379 	&	232.166 	&	210.435 	&	218.981 	&	198.963 	&	200.0   	\\
W	&	BCC	&	286.450 	&	267.864 	&	318.164 	&	306.480 	&	301.692 	&	296.696 	&	329.641 	 &	326.147 	&	357.702 	&	325.181 	&	340.697 	&	309.646 	&	323.2  	\\
Ir	&	FCC	&	301.726 	&	259.177 	&	359.776 	&	336.899 	&	329.294 	&	455.665 	&	373.098 	 &	370.375 	&	365.836 	&	355.695 	&	398.543 	&	344.228 	&	355.0  	\\
Pt	&	FCC	&	205.013 	&	171.114 	&	263.095 	&	242.600 	&	232.858 	&	341.923 	&	276.529 	 &	273.512 	&	267.401 	&	257.178 	&	302.235 	&	244.982 	&	278.3  	\\
Au	&	FCC	&	106.016 	&	89.683 	&	155.099 	&	139.849 	&	129.370 	&	191.343 	&	148.255 	&	 159.795 	&	156.150 	&	146.611 	&	189.275 	&	136.817 	&	173.2  	\\
Pb	&	FCC	&	34.303 	&	31.164 	&	44.442 	&	41.991 	&	39.015 	&	25.185 	&	42.843 	&	44.497 	&	41.383 	&	 40.978 	&	52.294 	&	39.735 	&	46.8$^{\ast}$	\\

				\hline
		\end{tabular}
}

\end{table*}

%

\begin{figure*}[ht!]
     \begin{center}
        \subfigure{%
            \label{fig3a}
            \includegraphics[width=0.47\textwidth]{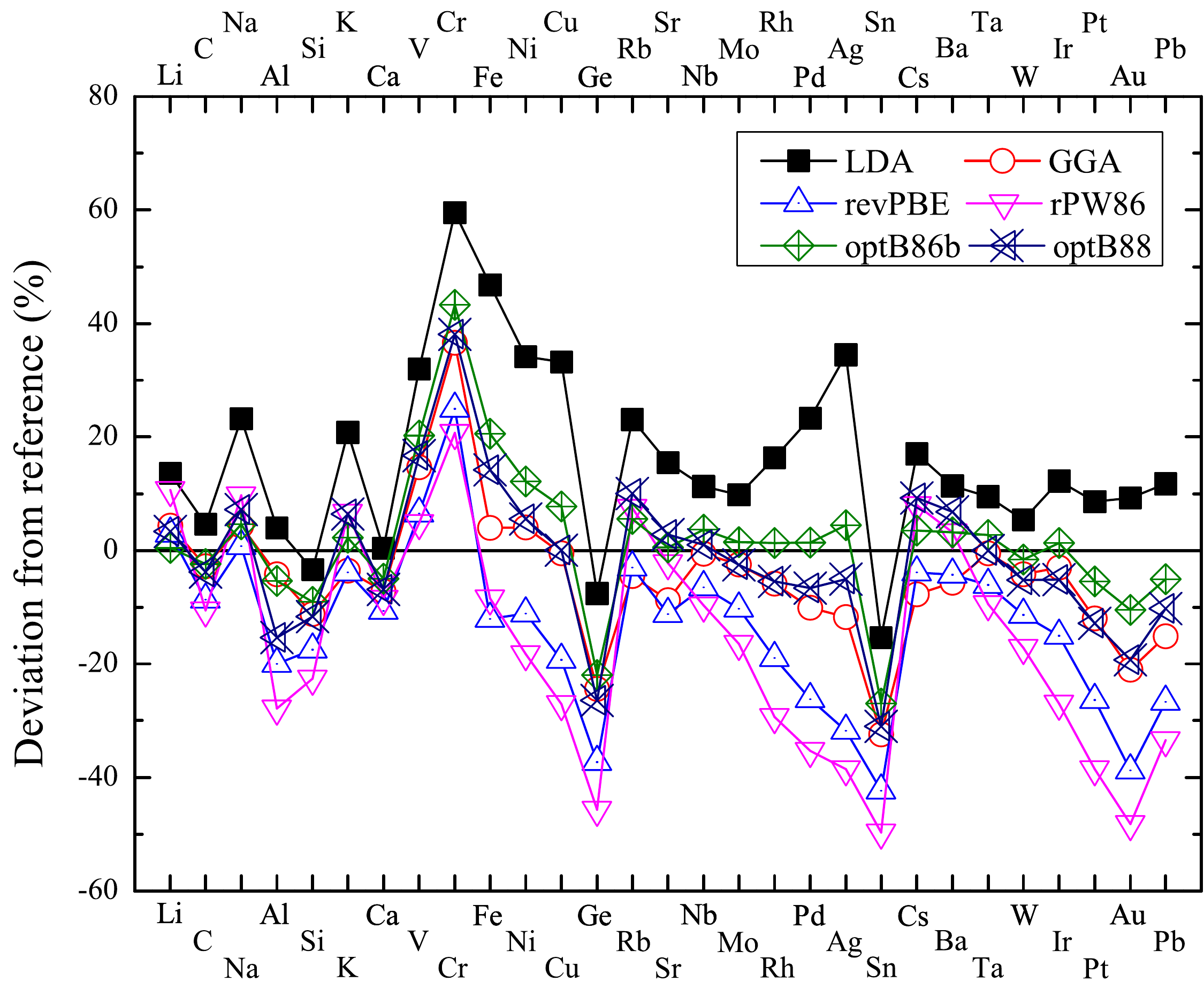}
        }%
        \subfigure{%
           \label{fig3b}
           \includegraphics[width=0.47\textwidth]{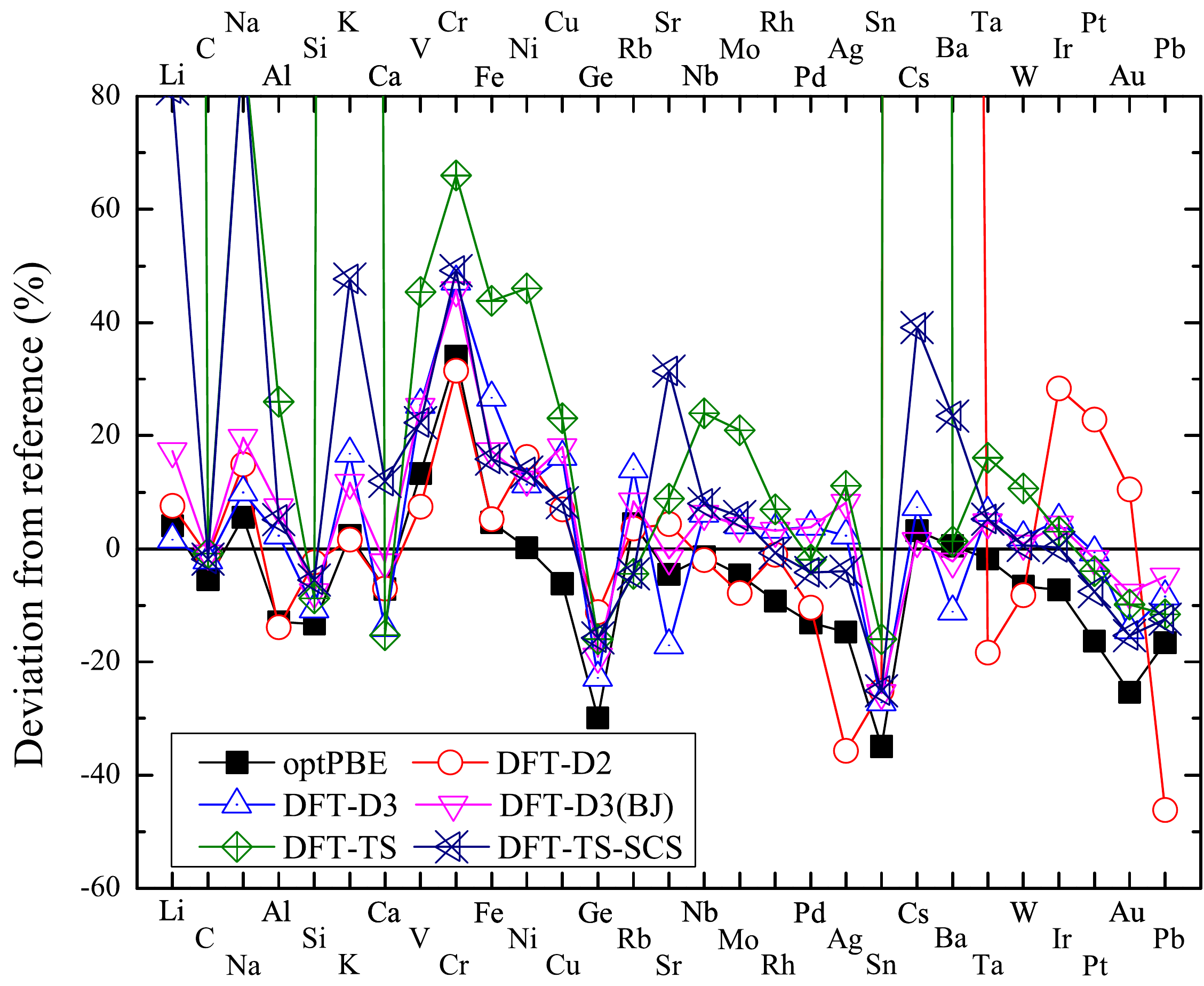}
        }
    \end{center}
    \caption{ Relative errors in the calculated bulk moduli with respect to the experimental values. Results are shown for the ten vdW functionals and the standard DFT functionals of LDA and GGA. The positive (negative) values in the relative errors represent the larger (smaller) bulk moduli than the experimental values.}
    \label{fig3}
\end{figure*}

The bulk moduli calculated with the ten vdW functionals are presented in Table \ref{table:BM}, and the relative errors in the calculated bulk moduli with respect to the experimental values are shown in Fig.~\ref{fig3}. In general, the trend of the bulk moduli is adequately reflected in the behavior of equilibrium lattice constants. Fig. \ref{fig3} clearly shows that the smaller equilibrium lattice constants lead to larger values of bulk moduli. As expected in the calculated equilibrium lattice constants, the DFT-TS and DFT-TS-SCS vdW functionals show worse results for alkali metals (Li, Na, K, Cs), and the DFT-D2 functional gives a relative error even over 100\%{} for Cs and Ba. In the case of the vdW-DF methods, for most of elements considered herein, the revPBE-vdW and rPW86-vdW2 give worse results than the other vdW-DF functionals (optB86b-vdW, optB88-vdW, and optPBE-vdW).

\begin{figure*}[ht!]
     \begin{center}
        \subfigure{%
            \label{fig4a}
            \includegraphics[width=0.47\textwidth]{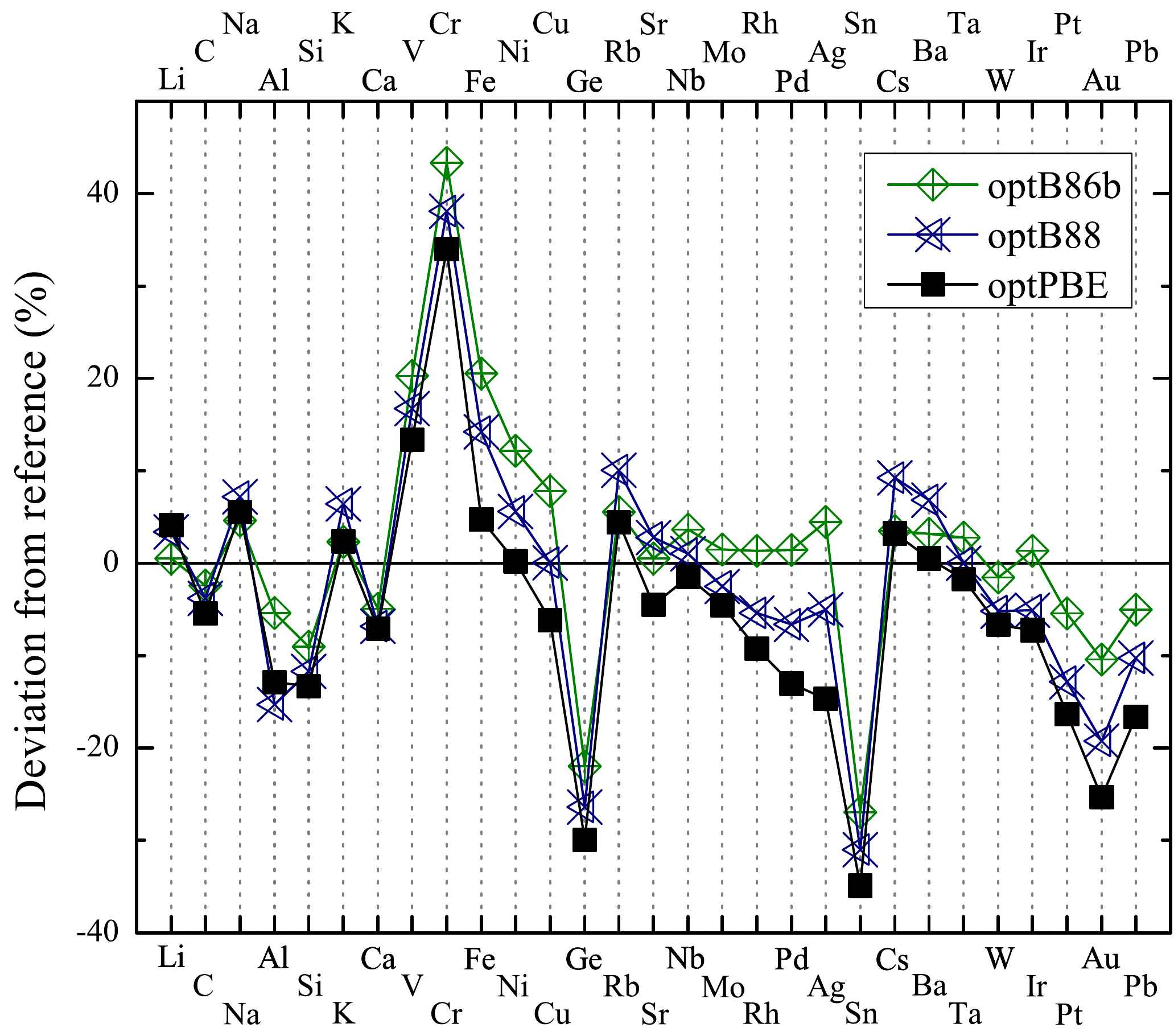}
        }%
        \subfigure{%
           \label{fig4b}
           \includegraphics[width=0.47\textwidth]{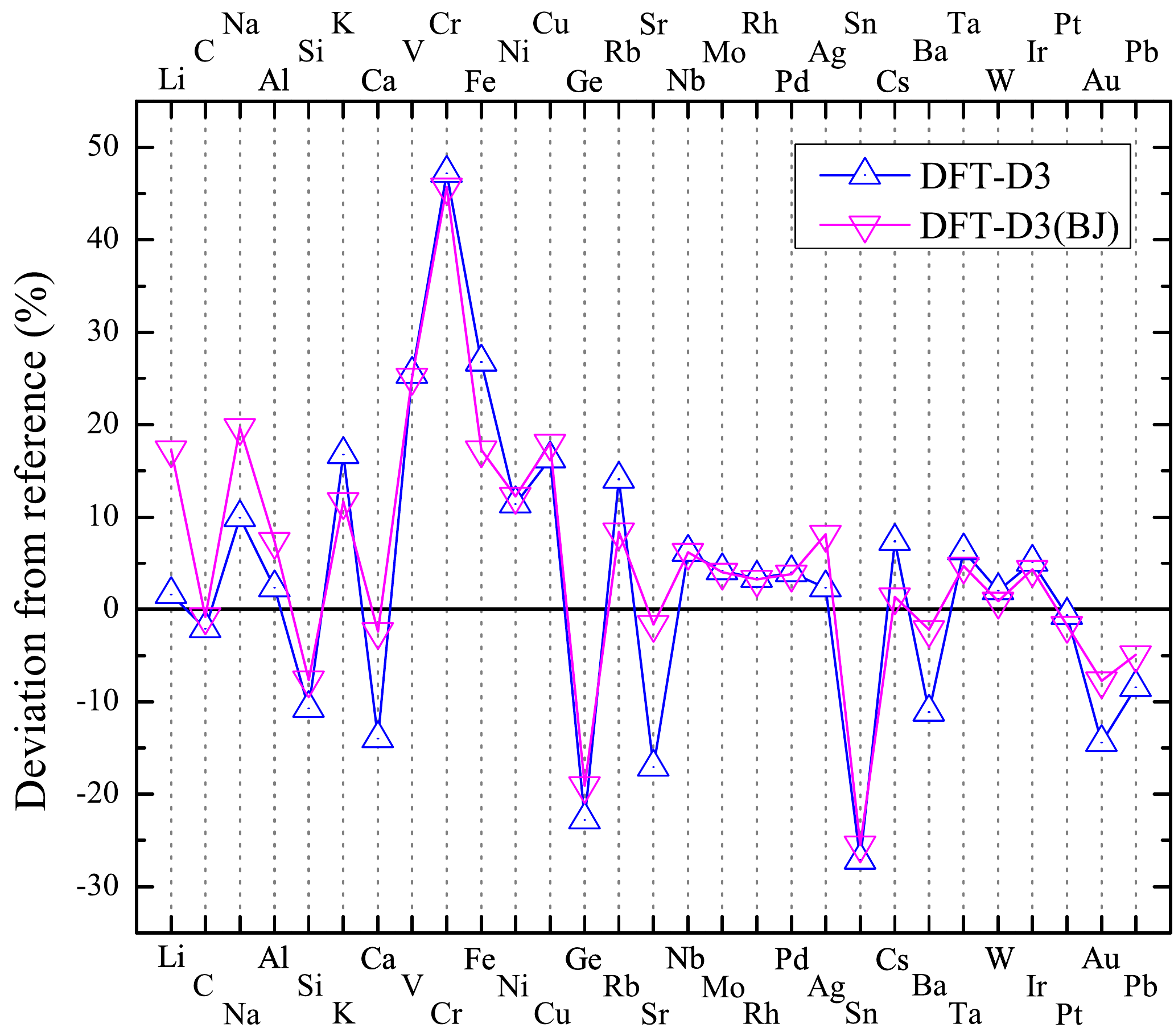}
        }
    \end{center}
    \caption{ Comparison of the relative errors in the calculated bulk moduli with respect to the experimental values. }
    \label{fig4}
\end{figure*}

 Next we discuss the differences among the five vdW results of optB86b-vdW, optB88-vdW, optPBE, DFT-D3, and DFT-D3(BJ). In the case of the DFT-D3 and DFT-D3(BJ) functionals, as expected from the calculated equilibrium lattice constants,
 the DFT-D3 functional shows either comparable or better results than the DFT-D3(BJ) functional (see Fig. \ref{fig4}). Better performance of DFT-D3 is observed for alkali and alkali-earth metals. In the case of the optB86b-vdW, optB88-vdW, and optPBE-vdW, the calculated results show very similar behavior, and the optB86b-vdW functional performs better for elements with large atomic number (see Fig. \ref{fig4}).


\subsection{Cohesive energy}
\label{CohesiveEnergy}
\begin{table*}
\caption{Cohesive energies (in eV/atom) for bulk solids with BCC, FCC, and DIA crystal structures using the ten vdW functionals. In addition, standard DFT calculation results using the LDA and the GGA are also shown. The theoretical values are compared to the experimental values. All the experimental cohesive energies \cite{kittel} were corrected by the zero-point vibration energy $E_{\rm ZPV}$ calculated using the Debye temperature \cite{csonka}. }
\label{ECOH}
\noindent \adjustbox{max width=\textwidth}{
		\begin{tabular}{ccccccccccccccc}
			\hline \hline
			Solid   & Crystal & revPBE & rPW86 & optB86b & optB88 & optPBE & DFT & DFT & DFT      & DFT  & DFT        & PAW & PAW & Expt. \\
			            & structure & vdW      & vdW2   & vdW        & vdW     & vdW      & D2   & D3    & D3(BJ)  & TS   & TS-SCS & LDA & GGA &  \\
			\hline

Li	&	BCC	&	1.540 	&	1.490 	&	1.640 	&	1.586 	&	1.622 	&	1.732 	&	1.707 	&	1.782 	&	3.873 	&	 2.625 	&	1.808 	&	1.605 	&	1.663 	\\
C	&	DIA	&	7.323 	&	7.208 	&	8.034 	&	7.891 	&	7.732 	&	8.065 	&	7.959 	&	8.043 	&	8.053 	&	 8.012 	&	8.998 	&	7.851 	&	7.586 	\\
Na	&	BCC	&	1.035 	&	0.929 	&	1.130 	&	1.070 	&	1.115 	&	1.225 	&	1.246 	&	1.249 	&	2.576 	&	 1.902 	&	1.258 	&	1.088 	&	1.128 	\\
Al	&	FCC	&	3.004 	&	2.596 	&	3.643 	&	3.379 	&	3.339 	&	3.800 	&	3.709 	&	3.888 	&	4.137 	&	 3.879 	&	4.052 	&	3.544 	&	3.431 	\\
Si	&	DIA	&	4.369 	&	4.167 	&	4.878 	&	4.744 	&	4.656 	&	4.839 	&	4.761 	&	4.926 	&	4.899 	&	 4.880 	&	5.344 	&	4.614 	&	4.693 	\\
K	&	BCC	&	0.851 	&	0.758 	&	0.935 	&	0.890 	&	0.926 	&	0.929 	&	0.997 	&	0.986 	&	4.196 	&	 1.438 	&	1.022 	&	0.869 	&	0.943 	\\
Ca	&	FCC	&	1.627 	&	1.402 	&	2.001 	&	1.875 	&	1.832 	&	2.086 	&	2.054 	&	2.137 	&	2.514 	&	 2.798 	&	2.204 	&	1.903 	&	1.862 	\\
V	&	BCC	&	5.075 	&	5.094 	&	5.969 	&	5.736 	&	5.407 	&	5.795 	&	5.954 	&	6.123 	&	6.613 	&	 6.113 	&	6.809 	&	5.413 	&	5.347 	\\
Cr	&	BCC	&	3.845 	&	3.906 	&	4.717 	&	4.483 	&	4.355 	&	4.471 	&	4.482 	&	4.559 	&	5.436 	&	 5.054 	&	5.728 	&	4.050 	&	4.161 	\\
Fe	&	BCC	&	4.521 	&	4.074 	&	5.560 	&	5.111 	&	5.065 	&	5.585 	&	5.474 	&	5.524 	&	6.075 	&	 5.628 	&	6.511 	&	5.161 	&	4.326 	\\
Ni	&	FCC	&	4.154 	&	4.050 	&	5.324 	&	4.978 	&	4.691 	&	5.275 	&	5.199 	&	5.268 	&	5.788 	&	 5.332 	&	6.063 	&	4.797 	&	4.484 	\\
Cu	&	FCC	&	2.975 	&	2.868 	&	3.755 	&	3.572 	&	3.399 	&	3.901 	&	3.994 	&	4.075 	&	4.092 	&	 3.840 	&	4.521 	&	3.485 	&	3.523 	\\
Ge	&	DIA	&	3.483 	&	3.444 	&	4.018 	&	3.921 	&	3.787 	&	4.061 	&	3.906 	&	4.025 	&	4.047 	&	 4.006 	&	4.611 	&	3.742 	&	3.881 	\\
Rb	&	BCC	&	0.778 	&	0.698 	&	0.862 	&	0.826 	&	0.853 	&	0.868 	&	0.902 	&	0.879 	&	0.775 	&	 0.775 	&	0.931 	&	0.774 	&	0.857 	\\
Sr	&	FCC	&	1.364 	&	1.130 	&	1.736 	&	1.607 	&	1.566 	&	1.873 	&	1.759 	&	1.810 	&	2.495 	&	 2.515 	&	1.894 	&	1.609 	&	1.734 	\\
Nb	&	BCC	&	6.797 	&	6.613 	&	7.767 	&	7.494 	&	7.345 	&	7.697 	&	7.617 	&	7.804 	&	8.458 	&	 8.322 	&	8.636 	&	7.056 	&	7.597 	\\
Mo	&	BCC	&	6.148 	&	5.966 	&	7.257 	&	6.954 	&	6.770 	&	8.411 	&	6.900 	&	8.487 	&	7.845 	&	 7.630 	&	8.310 	&	7.763 	&	6.864 	\\
Rh	&	FCC	&	5.400 	&	5.093 	&	6.657 	&	6.341 	&	6.057 	&	6.720 	&	6.598 	&	6.667 	&	6.818 	&	 6.528 	&	7.622 	&	6.021 	&	5.797 	\\
Pd	&	FCC	&	3.254 	&	3.209 	&	4.248 	&	4.038 	&	3.785 	&	4.371 	&	4.319 	&	4.399 	&	4.009 	&	 3.967 	&	5.058 	&	3.743 	&	3.917 	\\
Ag	&	FCC	&	2.216 	&	2.206 	&	2.960 	&	2.821 	&	2.625 	&	3.084 	&	3.008 	&	3.098 	&	2.966 	&	 2.876 	&	3.631 	&	2.519 	&	2.972 	\\
Sn 	&	DIA	&	3.025 	&	3.007 	&	3.512 	&	3.437 	&	3.307 	&	3.529 	&	3.378 	&	3.488 	&	3.515 	&	 3.428 	&	4.001 	&	3.199 	&	3.159 	\\
Cs	&	BCC	&	0.743 	&	0.679 	&	0.832 	&	0.656	&	0.820 	&	4.009 	&	0.824 	&	0.805 	&	5.375 	&	 1.148 	&	0.883 	&	0.715 	&	0.808 	\\
Ba	&	BCC	&	1.689 	&	1.520 	&	2.087 	&	1.995 	&	1.904 	&	4.881 	&	1.994 	&	2.079 	&	3.130 	&	 2.858 	&	2.248 	&	1.876 	&	1.911 	\\
Ta	&	BCC	&	7.645 	&	7.220 	&	8.878 	&	8.499 	&	8.285 	&	9.809 	&	9.008 	&	9.080 	&	9.562 	&	 9.424 	&	9.807 	&	8.411 	&	8.123 	\\
W	&	BCC	&	8.152 	&	8.002 	&	9.322 	&	9.013 	&	8.822 	&	10.067 	&	9.107 	&	9.182 	&	9.662 	&	 9.412 	&	10.381 	&	8.483 	&	8.939 	\\
Ir	&	FCC	&	6.489 	&	5.767 	&	8.091 	&	7.602 	&	7.309 	&	9.257 	&	7.990 	&	8.006 	&	7.838 	&	 7.639 	&	9.132 	&	7.282 	&	6.981 	\\
Pt	&	FCC	&	5.006 	&	4.729 	&	6.185 	&	5.900 	&	5.597 	&	7.373 	&	6.342 	&	6.342 	&	6.078 	&	 5.929 	&	7.112 	&	5.578 	&	5.863 	\\
Au	&	FCC	&	2.673 	&	2.619 	&	3.601 	&	3.403 	&	3.175 	&	4.523 	&	3.691 	&	3.679 	&	3.408 	&	 3.319 	&	4.309 	&	3.035 	&	3.826 	\\
Pb	&	FCC	&	2.866 	&	2.868 	&	3.410 	&	3.319 	&	3.180 	&	3.507 	&	3.219 	&	3.384 	&	3.288 	&	 3.259 	&	3.831 	&	2.984 	&	2.040 	\\

			\hline
		\end{tabular}
}
\end{table*}


\begin{figure*}[ht!]
     \begin{center}
        \subfigure{%
            \label{fig5a}
            \includegraphics[width=0.47\textwidth]{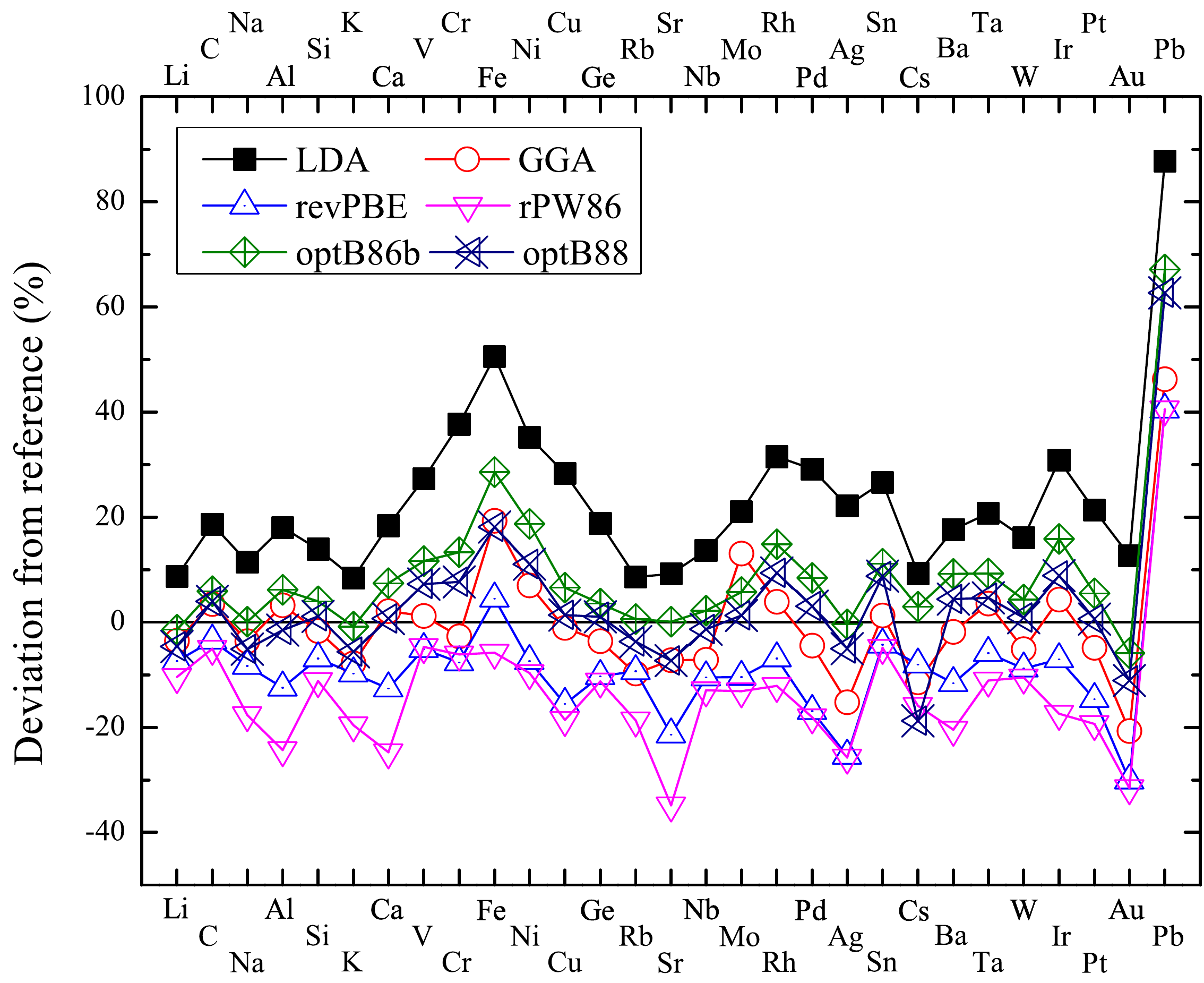}
        }%
        \subfigure{%
           \label{fig5b}
           \includegraphics[width=0.47\textwidth]{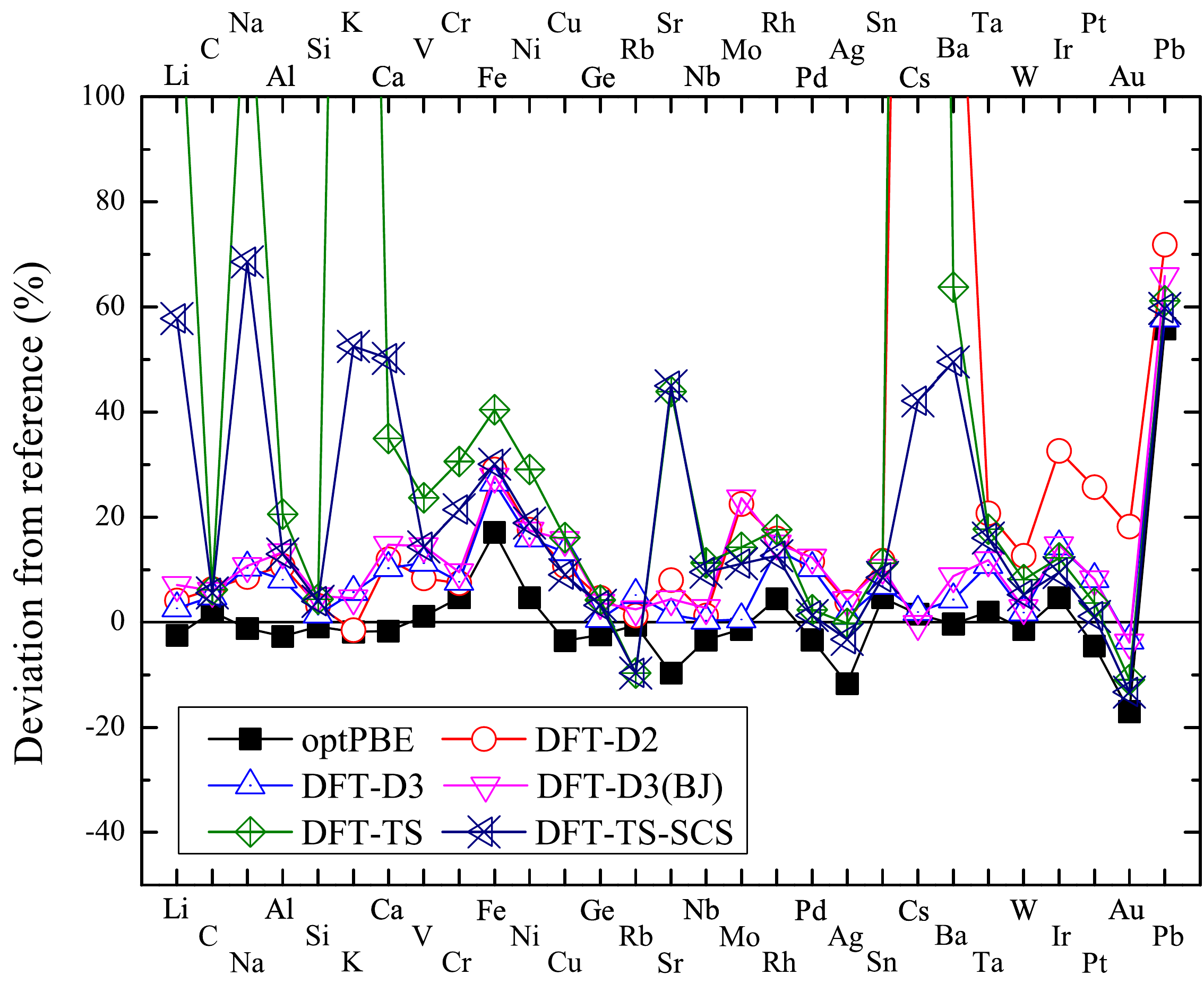}
        }
    \end{center}
    \caption{ Relative errors in the calculated cohesive energies with respect to the experimental values. Results are shown for the ten vdW functionals and the standard DFT functionals of LDA and GGA. The positive (negative) values in the relative errors represent the larger (smaller) cohesive energies than the experimental values.}
    \label{fig5}
\end{figure*}


\begin{figure*}[ht!]
     \begin{center}
        \subfigure{%
            \label{fig6a}
            \includegraphics[width=0.47\textwidth]{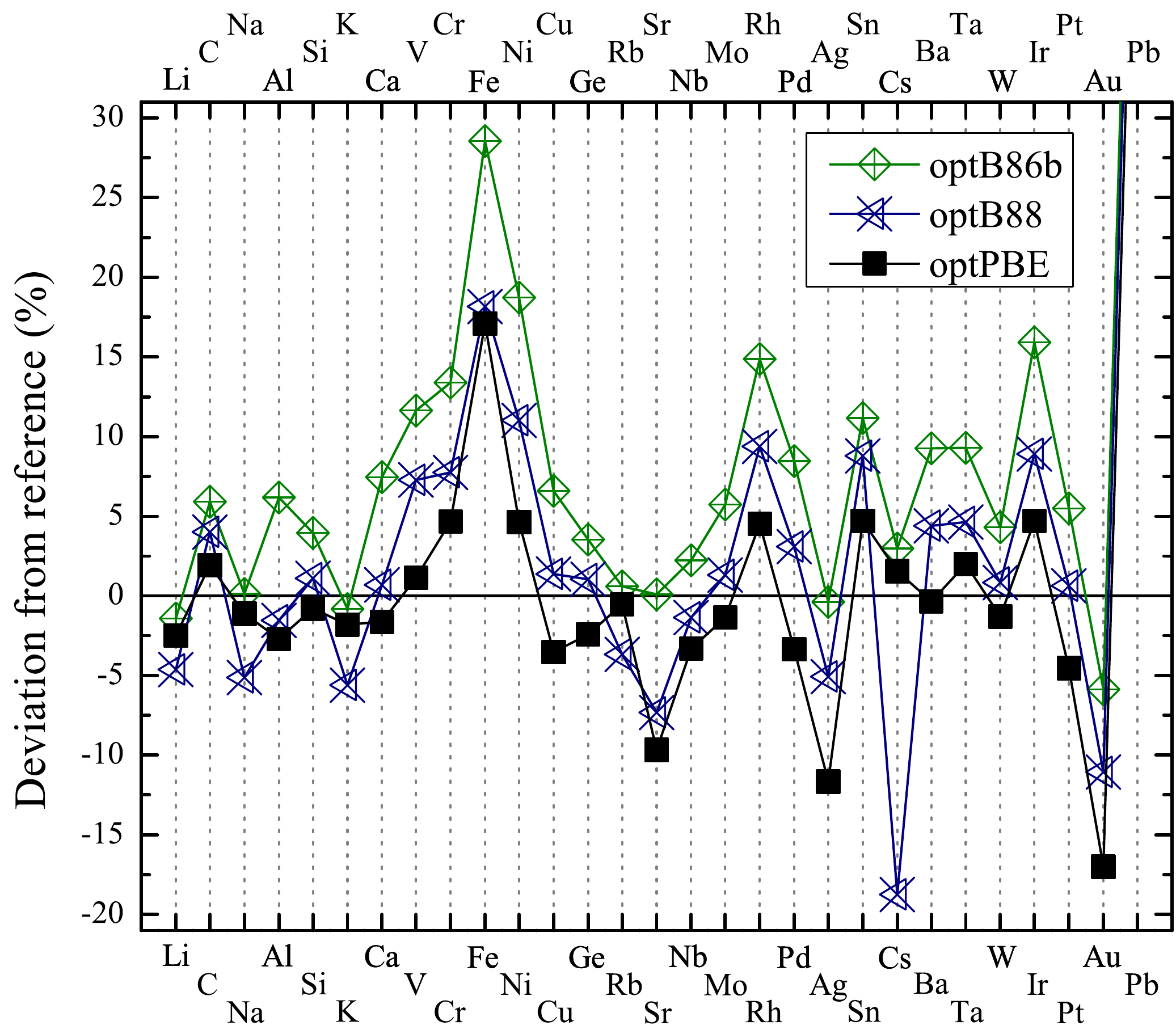}
        }%
        \subfigure{%
           \label{fig6b}
           \includegraphics[width=0.47\textwidth]{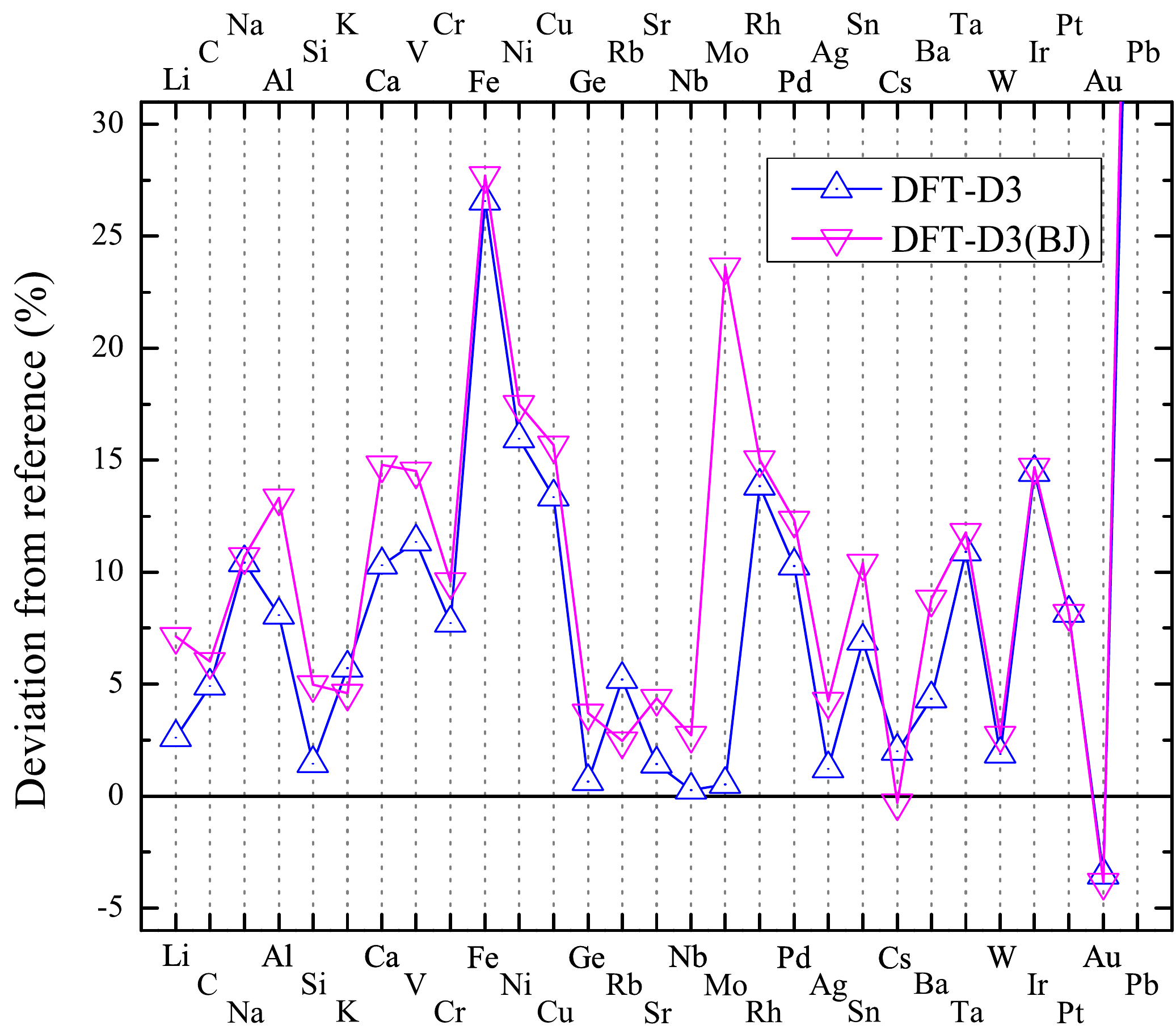}
        }
    \end{center}
    \caption{ Comparison of the relative errors in the calculated cohesive energies with respect to the experimental values. }
    \label{fig6}
\end{figure*}

The cohesive energies calculated with the ten vdW functionals are summarized in Table \ref{ECOH}, and the relative errors in the cohesive energies with respect to the experimental values are presented in Fig. \ref{fig5}. The cohesive energies $E_{\rm coh}$ are calculated using the following equation:
\begin{equation}
E_{\rm coh}=(E_{\rm tot}-n E_{\rm atom})/n,
\end{equation}
where $E_{\rm tot}$ and $E_{\rm atom}$ are the total energy of the system for atoms in the primitive unit cell at equilibrium and an isolated (free) spin-polarized atom, respectively, and $n$ is the number of atoms in the primitive unit cell. The experimental cohesive energies were corrected by the zero-point vibration energy $E_{\rm ZPV}$ calculated using the Debye temperature $\Theta_{\rm D}$, $E_{\rm ZPV} = (9/8) k_{\rm B} \Theta_{\rm D}$ \cite{csonka}.

For Pb, the relative errors in the calculated cohesive energies are as large as over 40\%{} for all the vdW functionals. In the case of the DFT-TS and DFT-TS-SCS functionals, poor performance is observed for alkali metals (Li, Na, K, Cs). The DFT-D2 functional shows poor performance for Cs and Ba. In the case of the vdW-DF functionals, the revPBE-vdW and rPW86-vdW2 functionals give lower cohesive energies than the other vdW-DF functionals.

Next we discuss the differences among the five vdW results of optB86b-vdW, optB88-vdW, optPBE, DFT-D3, and DFT-D3(BJ). The DFT-D3 functional shows either comparable or better performance compared to the DFT-D3(BJ) (see Fig. \ref{fig6}). In the case of the vdW-DF functionals, the optB86b-vdW, optB88-vdW, and optPBE-vdW functionals show very similar results, although the optB86b-vdW functional gives higher cohesive energies than the optB88-vdW and optPBE-vdW functionals (see Fig. \ref{fig6}).

\section{Summary}
\label{Summary}

In summary, we have investigated the lattice constants, the bulk moduli, and the cohesive energies for the bulk solids of 29 elements at equilibrium, using various vdW functionals based on the DFT in the VASP code. The assessed vdW functionals are classified into two groups. One is the vdW-DF functionals made by a proper choice of exchange functional, and the other is the vdW functionals of a dispersion-corrected DFT-D approach in which an atom-pairwise potential is added to a standard DFT result. The DFT-TS and DFT-TS-SCS functionals showed relatively poor performance for alkali and alkali-earth metals. Note that in the case of the DFT-TS and DFT-TS-SCS functionals, effective atomic volumes are used to calculate the dispersion coefficients. For the calculations, the partitioning of the electron density for each atom in a molecule or solid is performed and its result is then used to scale the dispersion coefficient with reference to the corresponding value for a free atom. Our calculations suggest that the partitioning of the electron density for effective atomic volumes may not be sufficiently accurate for delocalized alkali and alkali-earth metals. We obtained a general trend that the vdW functionals (optB86b-vdW, optB88-vdW, and optPBE-vdW) with optimized exchange functionals and the DFT-D vdW functionals [DFT-D3 and DFT-D3(BJ)] give better results than the original revPBE-vdW and rPW86-vdW2 functionals. To further aid in our understanding, we also discussed the differences among the vdW results of optB86b-vdW, optB88-vdW, optPBE-vdW, DFT-D3, and DFT-D3(BJ). These five vdW functionals showed very similar results. The DFT-D3 functional with zero damping gave either comparable or better performance compared to the DFT-D3(BJ) with damping. In the case of the vdW functionals with optimized exchange functionals, the optB86b-vdW showed slightly better performance in the equilibrium lattice constants and the bulk moduli compared to the other vdW functionals of optB88-vdW and optPBE-vdW. For the cohesive energies, the vdW functionals of optB86b-vdW, optB88-vdW, and optPBE-vdW functionals showed very similar results with smaller variation compared to the original vdW-DF methods of revPBE-vdW and rPW86-vdW2 and the standard LDA method. The results we present in this study provide fundamental information on how the various vdW functionals perform for the selected solid elements, including alkali, alkali-earth, and transition metals, with BCC, FCC, and diamond structures as the ground state structure.

\section{Acknowledgments}
This research was supported by Nano Material Technology Development Program (2012M3A7B4049888) through the National Research Foundation of Korea (NRF) funded by the Ministry of Science, ICT and Future Planning (MSIP), and Priority Research Center Program (2010-0020207) through NRF funded by the Ministry of Education (MOE). Calculations were performed by using the supercomputing resources (KSC-2014-C1-002) of the Korea Institute of Science and Technology Information (KISTI) and Korea Research Environment Open NETwork (KREONET), and the Partnership \& Leadership for the nationwide Supercomputing Infrastructure (PLSI).


\bibliographystyle{elsarticle-num}
\nocite{*}


\bibliography{reference}







\end{document}